\documentclass[preprint]{sig-alternate}
\pdfoutput=1
\usepackage{graphicx}
\usepackage{algorithmic}

\begin{document}

\title{Packet flow analysis in IP networks via abstract interpretation}

\numberofauthors{4}
\author{
\alignauthor Raghavan Komondoor\\
\affaddr{Indian Institute of Science, Bangalore}\\
\email{raghavan@csa.iisc.ernet.in}
\alignauthor K. Vasanta Lakshmi\\
\affaddr{Indian Institute of Science, Bangalore}\\
\email{kvasanta@csa.iisc.ernet.in}
\and
\alignauthor Deva Seetharam\\
\affaddr{IBM Research India}\\
\email{dseetharam@in.ibm.com}
\alignauthor Sudha Balodia\\
\affaddr{Indian Institute of Science, Bangalore}\\
\email{sudha.balodia@gmail.com}
}

\toappear{}
\maketitle

%% For describing the input network itself
\newcommand{\nodes}{\ensuremath{\mathbb{N}}}
\newcommand{\fnodes}{\ensuremath{\mathbb{F}}}
\newcommand{\znodes}{\ensuremath{\mathbb{Z}}}
\newcommand{\conxns}{\ensuremath{\mathbb{E}}}
\newcommand{\link}[4]{\ensuremath{({#1},{#2}) \rightarrow ({#3},{#4})}}
  % args are: node m, interface i_1, node n, interface i_2, such that
  % nodeof{i_1} = m, nodeof{i_2} = n, (i_1, i_2) \in \conxns
\newcommand{\zoneips}[1]{\ensuremath{\mathit{addr}_{#1}}} % arg is a zone
\newcommand{\nodeof}[1]{\ensuremath{\mathit{node}({#1})}} % arg is a interface
\newcommand{\dnat}[1]{\ensuremath{{#1}.\mathit{dnat}}} % arg is a firewall
\newcommand{\snat}[1]{\ensuremath{{#1}.\mathit{snat}}} % arg is a firewall
\newcommand{\filter}[1]{\ensuremath{{#1}.\mathit{filt}}} % arg is a firewall
\newcommand{\route}[1]{\ensuremath{{#1}.\mathit{rt}}} % arg is an firewall
\newcommand{\drop}{\small{DROP}}
\newcommand{\accept}{\small{ACCEPT}}

%% For describing the concrete as well as abstract domains
\newcommand{\concdomain}{\ensuremath{\mathit{Pk}}}
\newcommand{\pwidth}{\ensuremath{\mathit{pkSz}}} % width of concrete packets (in bits)
\newcommand{\fwidth}{\ensuremath{\mathit{fldSz}}} % maximum field width
\newcommand{\pfields}{\ensuremath{\mathit{nFlds}}} % # fields in a conc packet

\newcommand{\currf}{\ensuremath{\mathit{curr}}}
\newcommand{\origf}{\ensuremath{\mathit{orig}}}
\newcommand{\ifnatedf}{\ensuremath{\mathit{ifNated}}}
\newcommand{\grd}[1]{\ensuremath{{#1}.\mathit{grd}}} % arg is a rule
\newcommand{\action}[1]{\ensuremath{{#1}.\mathit{act}}} % arg is a rule
\newcommand{\natfield}[1]{\ensuremath{{#1}.\mathit{NAT\_field}}} % arg is a rule
\newcommand{\currpk}[1]{\ensuremath{{#1}.\currf}} % arg is an abs or conc packet
\newcommand{\origpk}[1]{\ensuremath{{#1}.\origf}} % arg is an abs or conc packet
\newcommand{\ifnated}[1]{\ensuremath{{#1}.\ifnatedf}} % arg is an abs packet

\newcommand{\absdomain}{\ensuremath{\mathit{AbsPk}}}
\newcommand{\packets}[1]{\ensuremath{\mathit{cp}_{#1}}} % arg is a node
\newcommand{\concr}[1]{\ensuremath{\mathit{conc}({#1})}} % arg is an
                                % abstract packet

%% Structures computed/used by the algorithm

\newcommand{\leaving}[1]{\ensuremath{{#1}.\mathit{from}}} % arg is a zone
\newcommand{\facts}[1]{\ensuremath{{#1}.\mathit{abs}}} % arg is a node
\newcommand{\forw}[1]{\ensuremath{\mathit{ff}_{#1}}}
   % arg is a rule or a table or a link
\newcommand{\join}{\sqcup}
\newcommand{\bigjoin}{\bigsqcup}
\newcommand{\backf}[1]{\ensuremath{\mathit{bf}_{#1}}}
   % arg is a rule or an interface 

%% General
\newcommand{\TODO}[1]{\textbf{#1}}
\newcommand{\tru}{\ensuremath{\mathit{true}}}
\newcommand{\fls}{\ensuremath{\mathit{false}}}

%% For describing the transfer functions
\newcommand{\filterruletf}[2]{\ensuremath{\mathit{filter\_ruleTF}({#1},{#2})}} % arg1 is a packet arg 2 is a rule
\newcommand{\filterchaintf}[2]{\ensuremath{\mathit{filter\_tableTF}({#1},{#2})}} 
\newcommand{\natruletf}[2]{\ensuremath{\mathit{nat\_ruleTF}({#1},{#2})}} 
\newcommand{\natchaintf}[2]{\ensuremath{\mathit{nat\_tableTF}({#1},{#2})}}

\newcommand{\natpacket}[2]{\ensuremath{\mathit{natPacket}({#1, #2})}} 
\newcommand{\updateoriginal}[2]{%
  \ensuremath{\mathit{update\_original\_packet}}%
  \ensuremath{({#1}, {#2})}}

\newcommand{\filterrules}{\ensuremath{\mathit{FilterRules}}}
\newcommand{\natrules}{\ensuremath{\mathit{NATRules}}}

\newcommand{\routinginfo}[1]{\ensuremath{\mathit{rt}({#1})}}  %arg is a interface

\newcommand{\reach}[1]{\ensuremath{{#1}.\mathit{reached\_packets}}} % arg is a rule
\newcommand{\dropp}[1]{\ensuremath{{#1}.\mathit{dropped\_packets}}} % arg is a rule

\begin{abstract}
Static analysis (aka offline analysis) of a model of an IP network is
useful for understanding, debugging, and verifying packet flow properties
of the network. There have been static analysis approaches proposed in the
literature for networks based on model checking as well as graph
reachability. Abstract interpretation is a method that has typically been
applied to static analysis of programs. We propose a new,
abstract-interpretation based approach for analysis of networks. We
formalize our approach, mention its correctness guarantee, and demonstrate
its flexibility in addressing multiple network-analysis problems that have
been previously solved via tailor-made approaches. Finally, we investigate
applications of our analysis for two novel problems -- automatically
generating test packets, and inferring a high-level policy for the network
-- which have been addressed in the past only in the restricted single-node
setting.

\end{abstract}

\section{Introduction}
\label{sec:intro}

Analysis of the flow of packets across an IP network is an important
problem. It has varied applications, such as identifying anomalies in
configuration files in routers~\cite{fireman-06},
testing of router implementations~\cite{al-shaer-segmentation-07}, checking
whether a network configuration satisfies a high-level policy of a network
administrator by querying properties of the
configuration~\cite{marmorstein-05,fang-journal-06}, and inferring such a
high-level policy automatically from the network
configuration~\cite{tongaonkar-07,horowitz-09}. However, such an analysis
is challenging, because packet routing in an IP network is a complex
activity. Routers intervene between subnets (i.e., fully connected
collections of hosts), and perform operations on packets such as filtering,
routing to adjacent routers or subnets, and transformation, e.g., for
network address translation (NAT). Each operation performed by a router is
predicated (i.e., guarded) by the current content of the header of the
packet, which, due to transformations, changes as the packet flows through
the network.  There are additional sources of complexity: The set of
operations performed by a router is not fixed once for all, but gets
modified as the network topology and load characteristics vary during
operation. Also, the outcome of some of these operations are dependent not
just on the content of the packet header, but also on the state of the
connection that the packet belongs to. All of this means that it is quite
difficult to analyze the flow of packets across the network.

The state-of-practice for analyzing reachability is to send test packets in
the actual network, using commercially available tools.  However, testing
does not give complete information about all possible packet flow outcomes,
because it is infeasible to send all possible packets across a
network. Several \emph{static} (or offline) analysis approaches,
e.g.,~\cite{xie-infocomm-05,fireman-06,al-shaer-reach-09}, have been
reported in the literature in order to overcome this disadvantage; these
approaches analyze a specification of the network topology and router
configurations (i.e., a \emph{model} of the network), and emit information
that over- or under-approximates all possible packet flows in the network.

\subsection{Contributions}

\emph{1)} Our primary contribution is an \emph{abstract
  interpretation}~\cite{Cousot77} based analysis for determining packet
flow properties in an IP network. To the best of our knowledge ours is the
first reported approach for this problem that is based on abstract
interpretation, which is a technique that has been typically applied to
analysis of properties of programs.  Abstract interpretation is a
customizable \emph{framework}, in the sense that it needs to be
instantiated with a \emph{lattice} (i.e., a domain of values to be used in
the analysis), and a set of \emph{transfer functions} operating on this
lattice. Therefore, the analysis designer has the flexibility to use
different lattices of differing precision for the same problem, and prove
that each one results in a semantically valid (but potentially approximate)
analysis wrt the most-precise analysis. We take advantage of this
capability by first spelling out a precise instantiation of our analysis,
which always terminates (because of bounded packet sizes), but which may be
expensive. Subsequently, we illustrate how to trade-off this precision for
scalability, while ensuring that the flow information we compute is an
over-approximation of the precise flows.  Previous static analysis
approaches for network analysis are hard-wired, and do not readily admit
such trade-offs within their overall approach.

\emph{2)} We show that abstract interpretation is a flexible framework,
capable of determining varying information about packet flows in a
network. The first variant of our analysis, discussed in
Section~\ref{sec:base-algorithm}, computes a formula for each intermediate
router that describes the set of packets that reach that
router. Determining reachability at intermediate nodes (i.e., routers) has
many applications, such as querying network policy
~\cite{marmorstein-05,fang-journal-06}, and identifying rule anomalies and
router mis-configurations~\cite{fireman-06}. The above-mentioned approaches
employ custom solutions, which miss certain packet flows (and hence may be
unsound) in the presence of cycles in the network. The problems addressed
by these approaches can be solved as straightforward postpasses after our
sound and generic reachability analysis.

The second variant of our analysis, discussed in
Section~\ref{sec:extended-algorithm}, computes information at each
intermediate router that not only represents the set of packets reaching
that router, but also the \emph{original} forms of these packets as when
they left their originating subnets (before they were transformed by
address translation along the way). 

\emph{3)} We propose a novel application of our analysis.
In previous work~\cite{tongaonkar-07,horowitz-09}
researchers have formulated the problem of
inferring a high-level policy of the
network, in the restricted setting of
single-router networks. We first generalize this
problem to the setting of a network of multiple routers, and then show how
to solve it using the second variant of our analysis.

%% TODO add some stuff in the beginning of kinds of network configuration
%% errors, their prevalence, and their effects

%% TODO we don't handle dynamic changes in network configuration, and
%% stateful routers.

%% TODO after rest of paper is ready go over intro and polish it again.
%% Especially the part about policy inference. Mention advantages of
%% inferring a policy.

%% Local Variables:
%% tex-main-file: "main.tex"
%% End:

\section{Related work}
\label{sec:related-work}

The previous static analysis techniques for IP networks that most closely
resemble ours are the ones based on transitive closure
analysis~\cite{xie-infocomm-05}, and graph propagation with bounded
unfolding of cycles~\cite{marmorstein-05,fang-journal-06,fireman-06}. All
of these approaches compute packet reachability information at all nodes in
the network. The work of Xie et al~\cite{xie-infocomm-05} is the seminal
work in the area of formally specified static analysis of networks. For
each pair of nodes $i, j$ in the network, they compute using Warshall's
transitive closure analysis a formula that represents the set of packets
leaving $i$ that eventually reach $j$ along all possible paths. Xie et al
pioneered the idea of uniformly treating filtering and NATing as
transformations on (representations of) sets of concrete packets. The other
approaches mentioned above, rather than using transitive closure, propagate
(representations of) sets of packets explicitly along the edges in the
network model. Our approach is similar to these approaches in this regard.

\begin{figure}
  \begin{center}
  \includegraphics[width=1.75in]{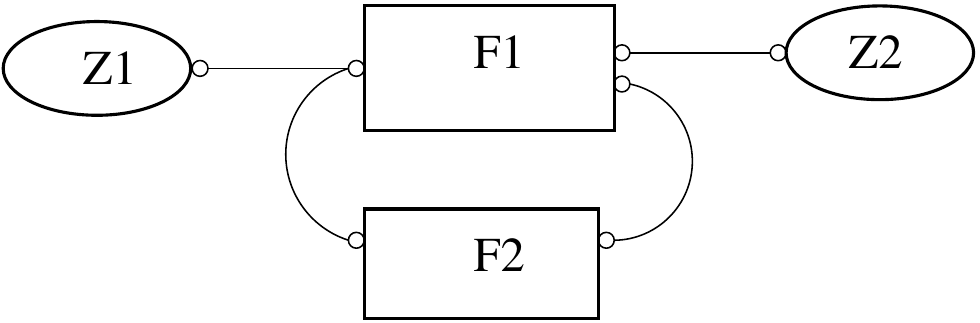} \\
  (a) \\
  \begin{small}
    \begin{itemize}
    \item F1 forwards all packets from Z1 or Z2 to F2 along the F1-F2 link
      on the right side.\\
    \item F2 filters out bad packets, SNATs src address field of good
      packets to a trusted address $T$, and forwards them to F1 along F2-F1
      link on left side.\\
    \item F1 forwards all packets that have src address $T$ (i.e., verified
      packets) to Z1 or Z2, based on their destination address.
    \end{itemize}
  \end{small}
  (b)
  \end{center}
  \caption{Packet propagation through cycles. (a) Example network (b)
    Routing configuration}
  \label{fig:cycle}
\end{figure}

The approaches mentioned above do not soundly analyze packet flows along
cyclic paths (i.e., they may miss certain packet flow). Consider the
example in Fig.~\ref{fig:cycle}. Part~(b) of this figure shows the
configuration rules in the firewalls F1 and F2, in plain English form for
the sake of clarity. Basically, F2 is a trusted subsidiary firewall that F1
sends all packets to for the sake of filtering. Therefore, e.g., a packet
from Z1 addressed to Z2 takes the following (cyclic) path:
Z1-F1-F2-F1-Z2. This example, although trivial, illustrates the
subsidiary-firewall idiom commonly employed by network administrators to
avoid overloading key firewalls (F1, in this case).  The cycle in this path
is not a ``useless'' cycle, in the sense that certain end-to-end flows can
happen only through this cycle. In general, for any integer $k$, it is
possible to construct a cycle going through $k$ routers such that certain
packets entering the cycle leave it only after going through the cycle $k$
times. Therefore, unrolling all loops a fixed number of times (which is the
idea behind the approached mentioned above) is not sufficient.  Abstract
interpretation involves an iterative analysis until a fix-point is reached,
and hence cleanly addresses this situation.

Model checking is another technique has been widely used in the
literature~\cite{matousek-08,jeffrey-09,al-shaer-reach-09} for static
analysis of networks. While the former two approaches model the flow of a
single packet through the network, Al Shaer's
approach~\cite{al-shaer-reach-09} models transitions of the set of all
packets in a network.  Since packet sizes in IP networks are bounded,
model-checking in this domain is capable of precise analysis even in the
presence of cycles.  Additionally, model checking can directly answer
general temporal properties, in additional to reachability (abstract
interpretation can answer restricted forms of temporal properties, too,
based on the abstraction chosen). Model checking, like abstract
interpretation, can also use abstract domains to compute approximate
solutions, e.g., as in the Slam~\cite{slam} approach (although the existing
model-checking based approaches for packet flow analysis do not do this).
The unique aspect of abstract interpretation is the formalism that
explicitly \emph{maps} the abstract values in the abstract lattice used to
concrete values in the concrete domain (e.g., sets of packets), and uses
this mapping as well as the properties of the given abstract transfer
functions to \emph{prove} that the analysis is sound (i.e., computes an
over-approximation of the precise information).

%% Local Variables:
%% tex-main-file: "main.tex"
%% End:

\section{Model and terminology}
\label{sec:notation}

A \emph{concrete packet} is an IP packet in a network. We only model the
headers of packets; let $\pwidth$ be the total number of bits in a packet
header, partitioned into $\pfields$ fields. We denote the fields of a
packet $p$ as $p.f_1, p.f_2, \ldots, p.f_p$. These fields include the
source address and port, and destination address and port. Let
$\concdomain$ represent the domain of all concrete packets.

We now describe our model of a network.  A network consists of a set of
nodes $\nodes$, which are partitioned into two categories: a set of
\emph{zones} (i.e. subnets) $\znodes$, which are terminal nodes, and a set
of \emph{firewalls} (i.e., routers) $\fnodes$, which are intermediate
nodes. We use zones to model organizational subnets as single units; i.e.,
we assume that each zone $z$ has a set of publicly visible IP addresses
$\zoneips{z}$ (with the sets of distinct zones being non-overlapping), and
that a packet leaving or entering a zone contains only public IP addresses
of that zone or other zones in its header.  We use $n, n_i$, etc., to
represent individual nodes, $z, z_i$, etc., to denote individual zones, and
$f, f_i$, etc., to denote individual firewalls. Each zone has a single
interface connected to the outside world, while each firewall has a set of
one or more interfaces. $\conxns$ is an irreflexive, symmetric, binary
relation on the set of all interfaces in the network, representing the
physical links between the interfaces; for any link $(i_1, i_2) \in
\conxns$, we assume that $i_1$ and $i_2$ do not belong to the same
firewall.
We use $\nodeof{i}$ to denote the zone or firewall to which
interface $i$ belongs. When we say $\link{m}{i_1}{n}{i_2}$, we mean $(i_1,
i_2) \in \conxns$, $\nodeof{i_1} = m$, and $\nodeof{i_2} = n$.

We now describe our model of how each firewall is configured; this is based
on the widely used package \texttt{Iptables}~\cite{iptables-tut}. 
Each firewall $f$ has four tables: a DNATing table $\dnat{f}$, a filtering
table $\filter{f}$, an SNATing table $\snat{f}$, and a routing table
$\route{f}$. Each packet entering a firewall through any of its interfaces
goes through the first three tables above, in the order mentioned, and
finally leaves through an interface as decided by the routing table. We
assume that firewalls are pure routers; i.e., they don't create or
ultimately accept packets. A filtering table is a sequence of filtering
rules, while each of the two NATing tables is a sequence of NATing
rules. Each rule $r$ (filtering or NATing) has two components: its
``guard'' $\grd{r}$, which is a propositional formula on the bits in a
packet header, and ``action'' $\action{r}$. A concrete packet $c$ is said
to \emph{match} a rule $r$ if $c$ satisfies the formula $\grd{p}$. A packet
entering a table is matched against each rule in the table sequentially
until a matching rule is found; the matching rule's action is then taken on
the packet, and the remaining rules are ignored.  For a filtering rule $r$
its action is either {\drop} or {\accept}; if a packet matches a filtering
rule $r$, it is thrown away if $\action{r}$ is {\drop}, and is sent out as
output from the table if $\action{r}$ is {\accept}. The final rule in any
filtering table has the guard {\tru} (i.e., is a \emph{default} rule). For
any NATing rule $f$, $\natfield{r}$ is a number which represents the field
in the packet header that is being NATed, while $\action{r}$ is a formula
representing a range of values.  If the NATing rule matches a packet $c$
then $c.\natfield{r}$ is overwritten with \emph{one} of the values in
$\action{r}$, and the hence transformed packet is sent out as output from
the table. If no rule in a NATing table matched a packet it is sent out
untransformed. DNATing rules write into the destination address or port
field, while SNATing rules write into the source address or port field.
The routing table $\route{f}$ of firewall $f$ is a function from the
interfaces in $f$ to formulas, each of which is
a constraint on destination addresses;
i.e., if a packet $c$, after having gone through the DNAT, filtering, and
SNAT tables in a firewall, has destination address $d$, it is then sent out
of one of the interfaces $i$ of $f$ such that $d$ satisfies the formula
$\route{f}(i)$.

Note in the discussion above that choices may have to be taken by NATing
rules as well as during the final routing step. We do not model how these
choices are made during network operation, and instead, in our analysis,
assume that \emph{all} choices are possible. Also, we assume the following
on the flow of concrete packets in the network: (a) There is no IP
spoofing; i.e., every packet leaving a zone $z$ has a source address that
matches $\zoneips{z}$, and a source port that is within the valid
port-range of $z$. (b) Every packet that enters the network from a zone
eventually reaches a zone $z$ that it is supposed to reach (i.e., its
destination address when it reaches $z$ matches $\zoneips{z}$), or gets
dropped by a filtering rule before it reaches any zone.

%% Neither of these assumptions are
%% central to the algorithm. They are common sense assumptions that most
%% networks enforce (or ought to enforce) anyway; our algorithm can be readily
%% modified to do away with these assumptions, but the results might not be
%% too useful, because ``junk'' packets would mix with ``useful'' packets and
%% pollute any analysis. \TODO{Our algorithm currently does not ignore useless
%%   cycles.}

\section{The base algorithm}
\label{sec:base-algorithm}

\begin{figure}
  \begin{small}
    \begin{center}
      \begin{minipage}{3.25in}
       1. $\currpk{p}$: Formula representing the set of concrete packets
       represented by $p$.\\
       2. $\origpk{p}$: Formula representing the set of original packets
       leaving a zone that, after flowing through the network, become the
       packets represented by \emph{curr}.\\
       3. $\ifnated{p}$: A vector of bits, one per field in a packet
       header. $\ifnated{p}.b_i$ is 1 means $\currpk{p}.f_i$ contains
       a value written by NATing (by some firewall).\\
       \textbf{Note:} The fields $\origf$ and $\ifnatedf$
       are used only by the second
       variant of our algorithm, discussed in
       Section~\ref{sec:extended-algorithm}.
      \end{minipage}      \\
       (a) \\[1em]
    
    \begin{minipage}{3.25in}
      \begin{algorithmic}[1]
        \STATE \emph{Inputs:} (1) A network configuration, (2) an
        originating zone $z_0$, (3) an \emph{abstract lattice}, whose
        elements are \emph{abstract values}, (4) an ``initial'' abstract
        value {\leaving{z_0}} at zone $z_0$, and (5) transfer functions for
        links.
    
        \STATE \emph{Outputs:} For each node $n$, an abstract value
        $\facts{n}$ (representing the set of concrete packets that could
        reach $n$).
        
        \STATE
        
        \STATE\label{ln:leaving-zone} Initialize $\facts{z_0}$ to
        $\leaving{z_0}$. Mark $z$.
        
        \STATE For all nodes $n$ other than $z$ initialize $\facts{n}$ to
        $\bot$ (the bottom element of the abstract lattice). 
        
        \WHILE{there exist marked nodes}

        \STATE Choose a marked node $m$, and unmark it.

        \FORALL{link $\link{m}{i_1}{n}{i_2}$}\label{ln:inner-loop-beg}

        \STATE\label{ln:call-xfer} Replace $\facts{n}$ with $\facts{n}
        \join \forw{(i_1,i_2)}(\facts{m})$.

        \STATE If node $n$ was unmarked, or if new value of $\facts{n}$
        different from old value, then mark $n$.
    
        \ENDFOR\label{ln:inner-loop-end}
        
        \ENDWHILE
    
      \end{algorithmic}
    \end{minipage} \\
    (b)
    \end{center}
  \end{small}
  \caption{(a) Fields in an abstract packet $p \in \absdomain$ (b)
    Propagation of abstract packets.}
\label{fig:algo}
\end{figure}

Instantiating an abstract interpretation requires us to specify
(a) an \emph{abstract} lattice, whose elements are called \emph{abstract
  values}, which is closed wrt the join operation (i.e., least upper bound,
or $\join$) (b) a directed graph on which the analysis is to be performed,
(c) \emph{transfer functions} for the edges in the graph, which specify the
abstract propagation semantics of the edges (as functions from abstract
values to abstract values), and (d) the \emph{initial} abstract value at
some designated \emph{originating} node $z_0$ of the graph. In our setting
the nodes in the network are the graph nodes, and each link
$\link{m}{i_1}{n}{i_2}$ in the network results in a graph edge $m
\rightarrow n$. We show the abstract interpretation algorithm in
Fig.~\ref{fig:algo}(b); this is basically Kildall's
algorithm~\cite{kildall}, instantiated to our setting. The idea behind the
algorithm is to keep track of an abstract value $\facts{m}$ at each node
$m$. In our setting, each abstract value is a set of \emph{abstract
  packets} from the domain $\absdomain$, where each abstract packet in turn
intuitively represents a set of concrete packets.  Whenever the abstract
value $\facts{m}$ at a node $m$ changes it is propagated through each
outgoing link $\link{m}{i_1}{n}{i_2}$ out of $m$ (see
lines~\ref{ln:inner-loop-beg}--\ref{ln:inner-loop-end}) using the transfer
function $\forw{(i_1,i_2)}$ of the link to the successor node $n$ of the
link; at $n$ this incoming value is \emph{joined} with the current abstract
value at $n$. The algorithm terminates when the abstract values at all
nodes stabilize (i.e., reach a fix-point); these values represent the
result of the algorithm.

Both variants of our algorithm share the basic structure mentioned
above. However, they differ in the content of the abstract packets, and in
the join operation and the transfer functions. We discuss the initial
variant of the algorithm in this section, and the second variant in the
next section. For the initial variant each abstract value is a
\emph{singleton} set, i.e., a single abstract packet.  Each abstract packet
$p$, in turn, is a structure with a single field $\currf$, which is a
propositional formula on the bits $b_0, b_1, \ldots, b_{\pwidth}$ in a
packet header; see Fig.~\ref{fig:algo}(a), ignoring the fields $\origf$
and $\ifnatedf$ for now (they are used by the second variant of our
algorithm). An abstract packet $p$ represents exactly the set of concrete
packets that satisfy the formula $\currpk{p}$. For instance, assuming
packet headers have only three bits, the formula $b_2 \wedge \neg b_0$
represents the set of packet headers $\{100, 110\}$.  This is formalized
using the mapping $\gamma$ from abstract packets to sets of concrete
packets, and its inverse mapping $\alpha$:

\begin{tabular}{lclp{2.5in}}
  $\gamma(p)$ &=& \multicolumn{2}{p{2.5in}}{$\{c \ |$ $c$ is a
    concrete packet, and $c$ satisfies $\currpk{p}\}$}  \\

  $\alpha(s)$ &= & $p$, & such that $\gamma(p) = s$
\end{tabular}

Since the abstract packet $\facts{n}$ at a node $n$ is meant to represent
the set of concrete packets that reach node $n$ along all possible paths,
it is natural for the join operator to be logical \emph{OR}; i.e., $p_1
\join p_2 = p_3$, where $\currpk{p_3} = \currpk{p_1} \vee \currpk{p_1}$.
The ``initial'' abstract value {\leaving{z_0}} at the originating zone
$z_0$ is an abstract packet such that its formula $\currf$ is satisfied by
all concrete packets whose source address is in $\zoneips{z_0}$.  The
transfer functions are shown in the appendix; ignore the statements labeled
``Variant 2'' or ``Inferring policy'' for now. Routine
{\filterchaintf{t}{\mathit{In}}} in Section~\ref{ssec:transf-filt-table} is
the pseudo-code for the transfer function for a filtering table $t$; 
(\emph{In} is the set of abstract packets coming into $t$, while the return
value is the set of abstract packets that come out of $t$. 
Similarly, {\natchaintf{t}{\_}} in Appendix~\ref{ssec:transf-nating-table} is
the pseudo-code for the transfer function for a NATing table $t$. For each
filtering or NATing table $t$ its transfer function $\forw{t}$ has the
signature $\absdomain \rightarrow \absdomain$, and captures the sequential
effect of all the rules in the table. For any abstract packet $p$, the
abstract packet $\forw{t}(p)$ represents precisely the set of concrete
packets that would result when the concrete packets represented by $p$ flow
through the table.

The routine {\filterruletf{r}{p}} in Appendix~\ref{ssec:transf-filt-rule}
is the pseudo-code for the transfer function for an individual filtering
rule $r$, applied to an incoming abstract packet $p$. Similarly, the
routine {\natruletf{r}{p}} in Appendix~\ref{ssec:transf-nating-rule} is for
the transfer function for an individual NATing rule $r$. In both these
transfer functions an incoming abstract packet could get split into two
outgoing abstract packets, one that matches the rule (and gets the formula
$\currpk{p} \wedge \grd{r}$), and one that does not match the rule (and
gets the formula $\currpk{p} \wedge \neg \grd{r}$).
In addition, each NATing rule $r$ updates the field indicated by
$\natfield{r}$ of the incoming packet $p$, by writing into this field the
values in the range $\action{r}$. This is accomplished by subroutine 
${\natpacket{p}{r}}$. 
The transfer function $\forw{(i_1,i_2)}$ of a link
$\link{m}{i_1}{n}{i_2}$ is shown in Appendix~\ref{ssec:transf-link}, and is
the only transfer function to be invoked directly by our propagation
algorithm in Fig.~\ref{fig:algo}. It works as follows: When given an
abstract packet $p$, it routes the packet through the tables $\dnat{m},
\filter{m}$, and $\snat{m}$, in that order. Finally, it refines the packet
to exclude concrete packets that it represents that have destination
addresses that do not satisfy the formula $\route{m}(i_1)$.

\begin{figure}[t]
  \begin{center}
  \includegraphics[width=2.7in]{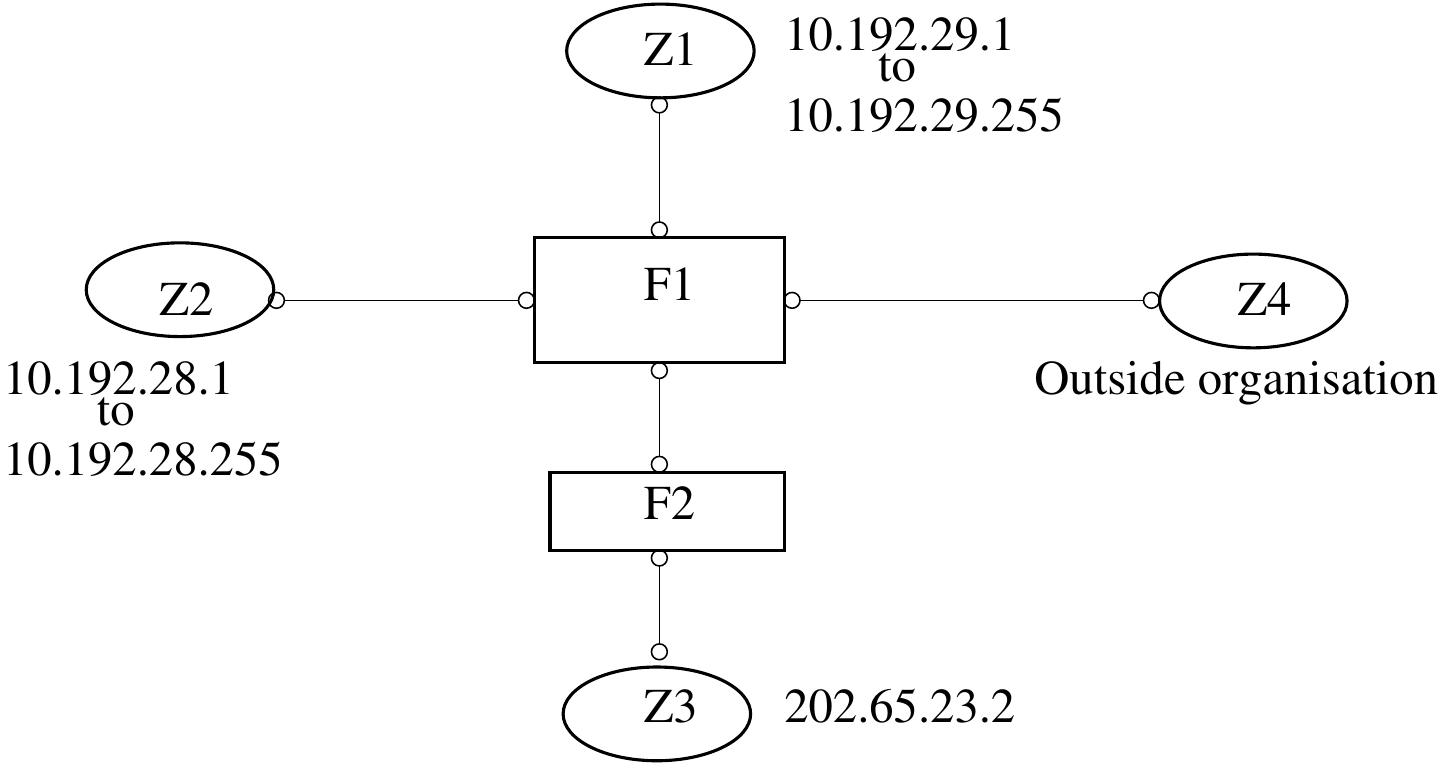} \\
  (a) \\
  \begin{scriptsize}
    \begin{tabular}{|l|l|c|}\hline
      \multicolumn{1}{|c}{\#} & \multicolumn{1}{|c}{\emph{Guard}} &
      \multicolumn{1}{|c|}{\emph{Action}} \\\hline
      \multicolumn{3}{|l|}{F1 filtering table:} \\\hline
      1. & $s$=10.192.29.[1-255], & DROP\\
         & $d$=209.85.153.85 & \\
      2. & $s$=10.192.28.[1-255], & DROP\\
         & $d$=209.85.153.85 & \\\hline
      \multicolumn{3}{|l|}{F1 SNAT table:} \\\hline
      3. & $s$=10.192.29.[1-255] & SNAT 202.67.34.[6-10] \\
      4. & $s$=10.192.28.[1-255] & SNAT 202.67.34.[1-5] \\\hline
      \multicolumn{3}{|l|}{F2 filtering table:} \\\hline
      5. & $s$=202.67.34.[6-10] & DROP \\\hline
    \end{tabular}
  \end{scriptsize}
  \\
  (b) \\
  \begin{scriptsize}
  \begin{minipage}{3.25in}
  \begin{tabular}{ll}
    $\facts{Z1}$ = & $<[$ 10.192.29.1-255 : {\tru}$]$, 
      $[$10.192.29.1-255 : {\tru}$]>$ \\ %% at z2

    $\facts{Z2}$ = & $<[$202.67.34.6-10 : 10.192.28.1-255$]$, \\
      & \ \ $[$10.192.29.1-255 : 10.192.28.1-255$]>$ \\ %% at z2

    $\facts{Z4}$ = &
    $<[$202.67.34.6-10 : $\neg\{$10.192.28.1-255, 10.192.29.1-255, \\
    & \ \    209.85.153.85, 202.65.23.2$\}]$, \\
    
    &  \ \ $[$10.192.29.1-255 : $\neg\{$10.192.28.1-255, 10.192.29.1-255, \\
      
    &  \ \   209.85.153.85, 202.65.23.2$\}]>$ \\ %% z4
    \multicolumn{2}{c}{(c)}
  \end{tabular}
  \end{minipage}
  \end{scriptsize} \\
  (c)
  \end{center}
  \caption{(a) Example network (b)
    Firewalls configuration (c) Reached abstract packets, with Z1 as origin}
\label{fig:main-example}
\end{figure}

For an illustration consider the example network in
Fig.~\ref{fig:main-example}(a), the firewall configurations of which are
shown in part~(b). Zones Z1, Z2, and Z3 belong to the same organization,
while Z4 models the outside internet. Firewall F1 is the primary gateway of
the organization. Zones Z1 and Z2 use private IP addresses; Z1 uses the
private address range 10.192.29.[1-255], while Z2 uses the private address
range 10.192.28.[1-255]. F1 drops packets from these two zones to the
blocked outside host 209.85.153.85 -- see Rules~1 and~2 (in the guards $s$
stands for source address and $d$ stands for destination address). Zone Z3
provides a service to the rest of the organization, and is accessible at
the (public) address 202.65.23.2 (i.e., it is outside the organization's
intranet). F1's SNAT table translates the source addresses of packets
coming from zone Z1 to the (public) range 202.67.34.[6-10] (see Rule~3),
and the source adresses of packets coming from zone Z2 to the (public)
range 202.67.34.[1-5] (see Rule~4).  Finally F2 denies access to Z3 for
packets whose source address is in the range 202.67.34.[6-10] (see
Rule~5). This range corresponds to packets that came originally from Z1 and
were NATed by F1.

Consider a run of our algorithm starting from zone Z1. No abstract packet
reaches zone Z3, because F2 denies access from Z1. The abstract packet
reaching each other zone is shown in Fig.~\ref{fig:main-example}(c). Our
notation is as follows: The text inside each pair of angled brackets is an
abstract packet. There are two components inside each abstract packet $p$,
delimited by square brackets. The first component is $\currpk{p}$; ignore
the second component for now. For convenience, we denote the formula
$\currpk{p}$ as a pair of constraints on the source and destination fields,
respectively, separated by a colon. 

\subsection{Correctness and complexity}

The abstract interpretation framework guarantees termination and
correctness as long as the instantiation (i.e., the lattice, and the
transfer functions) satisfy certain sufficient conditions (we refer you to
Cousot and Cousot's paper~\cite{Cousot77} for the details of the sufficient
conditions). Since the formula $\currpk{\facts{n}}$ at any node $n$ keeps
monotonically getting weaker (due to joins), and since the number of
distinct formulas is finite (due to the fixed packet width), the algorithm
is guaranteed to reach a fix point and terminate.

Our transfer functions are precise, in the sense described earlier. Also,
our abstract lattice is precise, in the sense that for any set $s$ of
concrete packets, $\gamma(\alpha(s)) = s$ (the abstract lattice is called
imprecise if for any set $s$ we have $\gamma(\alpha(s)) \supset
s$). Therefore, our analysis is precise; i.e., the final abstract packet
$\facts{n}$ at each node $n$ represents precisely the set of concrete
packets that will eventually flow through $n$ (after passing through all
its three tables) assuming an initial configuration wherein all concrete
packets represented by the abstract packet $\leaving{z_0}$ start out from
zone $z_0$.

Reachability analysis in networks is an NP-complete
problem~\cite{jeffrey-09} (on the packet size). In the worst case there
could be an exponential number of paths to a node $n$ in a network and the
abstract packet $\currpk{\facts{n}}$ at this node could in the worst-case
be updated $O(2^{\pwidth})$ times during a run of the algorithm.  Our
precise abstract-interpretation formulation described so far, therefore,
will have similar  running time requirement as model checking
approaches reported in the literature~\cite{matousek-08,jeffrey-09}, which
also answer reachability.

\subsection{Precision-efficiency trade-offs}
\label{ssec:prec-effic-trade}

A key benefit of abstract interpretation is that it uses a join operation
to merge abstract values reaching any node; therefore, it is possible to
tweak the abstract packet structure, as well as the join operation to
improve efficiency (by reducing precision). We illustrate this idea by
considering one such optimization. Rather than have a single formula
describing all the $\currpk{p}$ bits in the packet header, we model an
abstract packet $p$ as a sequence of formulas $\currf_1, \currf_2, \ldots,
\currf_{\pfields}$, where $\pfields$ is the number of fields in a packet
header. This is typically called an \emph{independent attribute} analysis
in the program analysis literature, as opposed to a \emph{relational}
(i.e., precise) analysis. \emph{AND} and \emph{OR} operations are now done
separately on each pair of formulas (of corresponding fields), while
\emph{NOT} of any sequence of formulas is approximated as {\tru}
(otherwise, the negation of an abstract packet could result in exponential
number of abstract packets). Therefore, the worst-case number of updates to
the abstract packet at any node during a run of the algorithm is now
$O(\pfields * 2^{\fwidth})$, where $\fwidth$ is the number of bits in the
largest field. This is exponential on the size of the longest individual
field, as opposed to being an exponential on the total size of the packet,
which is a significant gain in practice. While this analysis may
over-approximate the packet flows in the network, it still has value; e.g.,
if it says that a certain (undesirable) packet flow is \emph{not} possible,
this is guaranteed to be the case. Also, one could start with an imprecise
analysis, and then progressively improve its precision using the idea of
counter-example guided abstraction refinement, e.g., as in
Slam~\cite{slam}, until the undesirable packet flow to be verified is
proved with certainty to be either possible or impossible.

%% TODO Spell out some implementation tips. For instance, BDDs could be
%%  used.

%% TODO Changes required to Vasanta's pseudo-code:
%%
%% Filtering:
%% - Separate out transfer function for a single filtering rule. Let its
%%   signature be Abs -> 2^Abs
%% - Get rid of ``M'' bit
%% - In transfer function for filtering chain, swap the two loops. That is,
%%   let the outer loop iterate over the rules in the chain, and let the
%%   inner loop iterate over the abstract packets that have reached so
%%   far. Let the transfer function of the whole chain also be Abs ->
%%   2^Abs.
%% - MISTAKE: You cannot add a packet in P.original as a whole to
%%   reached_packets. The packet that you had to reached_packets need to be
%%   constructed from P.original as well P.curr. The NATed fields should
%%   come from O.original while the non-NATed fields should come from
%%   P.curr.  Otherwise you lose precision.
%%
%% NATing:
%%
%% - As above, separate out the rule for a single rule.
%% - MISTAKE: P1.curr.p_i needs to be updated *even if* P1.ifNated.n_i is
%%   1.
%% - Don't put a packet into the finalQueue as soon as it matches one
%%   rule.
%% - MISTAKE: What do you with P2? Also, why is the second ``if'' statement
%%  (involving \neg rule.guard) nested inside the first ``if'' statement? 

%% Local Variables:
%% tex-main-file: "main.tex"
%% End:

\section{Extended algorithm}
\label{sec:extended-algorithm}

In the first variant of our algorithm, discussed in the previous section,
the field $\currpk{p}$ of any abstract packet $p$ represents the set of
packets that have reached the node where $p$ resides. Note that due to
NATing, the current form of these packets (as represented by \currpk{p})
could be different from their \emph{original} form when they originally
left the designated source zone $z_0$. In this variant of the algorithm we
extend the abstract packet to have another field $\origpk{p}$, which
represents the original forms of the packets represented by $\currpk{p}$
when they left $z_0$. This information, which basically augments the
reachability information, is likely to be useful in a variety of bug
detection, understanding, and verification tasks. We
explore a specific application of this analysis later in this section.  

In this variant an abstract value (i.e., abstract lattice element) is a
possibly non-singleton set of abstract packets. The ``initial'' abstract
value $\leaving{z_0}$ leaving the zone $z_0$ is a
singleton set containing an abstract packet $p$ whose $\currf$ and $\origf$
formulas are identical, and are satisfied by all concrete packets whose
source address is in $\zoneips{z_0}$.

As before, we formalize the semantics of each abstract packet by defining
$\alpha, \gamma$ maps that relate abstract packets to concrete packets. To
enable this we first extend our model of the concrete packets. We
let each concrete packet $c$ have two fields $\currpk{c}$ and $\origpk{c}$,
the first one of which represents its current contents, and keeps changing
as the packet flows through NATing rules, while the second one is fixed,
retaining its original form throughout. Now:

\begin{tabular}{lp{2.5in}}
  $\gamma(P)$ =& $\bigcup_{p \in P} \gamma(p)$, where $P$ is a set of
    abstr. packets\\

  $\gamma(p)$ =& $\{c \; |$ $c$ is a conc. packet,
    $\currpk{c}$ satisfies $\currpk{p}$,
    $\origpk{c}$ satisfies $\origpk{p}\}$, \\
       & where $p$ is an abstract packet \\

  $\alpha(C)$ =& $P$, such that $\gamma(P) = C$, where $C$ is a set
    of concrete    packets. 
\end{tabular}

In other words, the correctness guarantee of the algorithm is that if an
abstract packet $p$ is in the set $\facts{n}$ at some node $n$, then for
every concrete packet $c_1$ that satisfies the formula $\origpk{p}$ and for
every concrete packet $c_2$ that satisfies the formula $\currpk{p}$ there
is a path in the network from $z_0$  to
$n$ such that $c_1$ is in $\leaving{z_0}$ and $c_1$ becomes transformed to
$c_2$ by the time it reaches $n$ along the path.

In this setting
a precise way to define join of two sets of abstract packets $P_1$ and
$P_2$ is set union. However, we present an optimized version of this join in
Fig.~\ref{fig:2nd-variant-join} which is still precise, and is sufficient
to guarantee the correctness property mentioned above.

\begin{figure}
  \begin{small}
\begin{algorithmic}
  \STATE \emph{Input:} Two sets of abstract packets $P_1$ and $P_2$
  \STATE \emph{Result} = $\phi$
  \FORALL{abstract packets $p_1 \in P_1$}
    \STATE Let $\mathit{newCurr}$ = $\currpk{p_1}$ $\vee$
    $\bigvee_{\{p_2 \in P_2  |  \origpk{p_2} =
      \origpk{p_1}\}} \currpk{p_2}$
    \STATE \emph{Result} = \emph{Result} $\cup\ p$, where $p$ is a new
    abstract packet such that
    $\currpk{p} = \mathit{newCurr}$ and $\origpk{p} = \origpk{p_1}$
  \ENDFOR
  \RETURN \emph{Result}
\end{algorithmic}
  \end{small}
\caption{Optimized join operation}
\label{fig:2nd-variant-join}
\end{figure}

The transfer functions for this new lattice are the same ones discussed
earlier (shown in the appendix), except that the lines labeled ``Variant
2'' are now included; ignore the lines labeled ``Inferring policy'' for
now. The changes to transfer functions $\filterruletf{\_}{\_}$ and
$\natruletf{\_}{\_}$ can be summarized as follows: as each packet flows
through a rule, the ``orig'' version is refined using the guard of the
rule, but only updating the fields that have not been NATed yet. We keep
track of which fields in $p$ have been NATed so far by any firewall along the
path along which $p$ flowed, using
an auxiliary bit $\ifnated{p}.b_l$ for each field $l$ in the packet
header. The ``orig'' formula is not refined for fields that have been
NATed because for a NATed field the rule refers
to the \emph{new} (NATed) value, and not the original value.

\sloppypar
There are additional changes in the transfer function
$\natruletf{r}{p}$. If the field $\natfield{r}$ of $p$ is being NATed for
the first time in the history of this packet, we first \emph{extract} the
content of field $l$ from $\currpk{p}$ (which still represents the original
value of this field when the packet left its source zone) and \emph{copy}
it to the corresponding field in $\origpk{p}$. This is done by calling
routine {\updateoriginal{\_}{\_}}. We then update the field $\natfield{r}$
in $\currpk{p}$ by calling the {\natpacket{\_}{\_}} (this is the same as in
Variant 1 of the algorithm).

The transfer function described above is precise, in the sense that for any
abstract packet $p$ and any NATing table $t$, the abstract packet
$\forw{t}(p)$ represents precisely the set of concrete packets that would
result when the concrete packets represented by $p$ flow through the table.
The net result of this is that for any abstract packet $p$ at a node $n$,
$\origpk{p}$ precisely captures the original forms of the packets leaving
$z_0$ that reach $n$ and that are represented by $\currpk{p}$. 

The semantics of the copying of field $l$ from $\currpk{p}$ to $\origpk{p}$,
mentioned above, can be stated more precisely is as follows.  We extract
the original content of this field from $\currpk{p}$ as a formula $m_l$,
which represents the set $s_l$ of original concrete bit sequences that
reside in field $l$ of concrete packets represented by $p$ before the
NATing happens.  We then update the formula $\origpk{p}$, such that all
concrete packets represented by it now have a bit sequence from $s_l$ in
their $l$ field, but whose other fields are undisturbed.

%% A note about the transfer functions: the functions shown in the appendix
%% map an abstract packet to another abstract packet. In the current variant
%% of the algorithm an abstract value is a \emph{set} of abstract
%% packets. This is not an issue, however, because of the distributivity of
%% these functions over set union. Therefore, in line~\ref{ln:call-xfer} in
%% Fig.~\ref{fig:algo} we simply need to call the function $\forw{(i_1,i_2)}$
%% multiple times, once for each abstract packet in $\facts{m}$, and join the
%% results.

Consider again the example in Fig.~\ref{fig:main-example}, where we run the
analysis starting from zone Z1. Note that a single abstract packet $p$
(delimited by angle brackets) reaches zone Z4 (see Part~(c) of the
figure). The first component inside this abstract packet denotes
$\currpk{p}$; note that its source address is the address range
202.67.34.6-10 that was written by the NAT rule in F1. The second component
denotes $\origpk{p}$; note that its source address is the original source
address range  of the packet leaving Z1 (i.e., 10.192.29.1-255).

\subsection{Application: Inferring a high-level policy of a network}
\label{sec:appl-inferr-high}

Real-life networks can be large, with 5-500 intermediate
routers~\cite{xie-infocomm-05}. Configuring these routers correctly is a
complex and error-prone task.  In a study of 37 real firewalls
Wool~\cite{wool-study-04} found that each one of them was misconfigured,
and had security vulnerabilities. Therefore, it is important for network
administrators to have access to tools that infer a compact, high-level
policy from a network that has already been setup, to help them debug and
validate the configuration.  Tongaonkar et al~\cite{tongaonkar-07} and
Horowitz et al~\cite{horowitz-09} have proposed inferring a policy for a
\emph{single} firewall. In both these approaches the initial step is to find the
rules that have overlapping guards, and then to present a transformed, or
differently organized version of the ruleset. While Tongaonkar et al
\emph{flatten} the ruleset, by eliminating all overlap between them,
Horowitz et al organize the rules hierarchically, with rules with weaker
guards placed ``above'' rules with stronger guards. These ideas do not
extend cleanly to the setting of multiple firewalls connected as a
network. Due to the large number of rules in real networks, and because
different sets of rules may be correlated along different paths in a
network, it is not clear that rule correlations can be presented in a natural,
compact manner in this setting.

Our hypothesis is that in many cases it would help the administrator if for
each zone $z$, they are simply given an ``accept'' formula that
characterizes the set of packet headers that leave $z$ that eventually reach
\emph{some} other zone, and a ``reject'' formula that characterizes the set
of packet headers leaving $z$ that get dropped by some rule.  The two sets
may, in general, be overlapping; a non-empty overlap should be a matter of concern
to the administrator, because packets matching both these formulas may
reach some zone, or none at all, depending on the (non-deterministic)
route they take through the network.  This pair of formulas for zone $z$
is a \emph{high-level} policy,
in the sense that it is compact, and conveys useful end-to-end information
whose representation is not tied to the actual way in the which the network
configuration has been set up.

The first step in determining this high-level policy is to run our analysis
treating $z$ as the ``originating'' zone $z_0$. Then, the ``accept''
formula for zone $z$ is simply

\[\bigvee_{z_i \in \znodes-\{z\}} \origpk{\facts{z_i}}\]

If the set of all filtering rules in the network with $\drop$ as the action
is represented by $D$ then the ``drop'' formula for $z$ is

\[ \bigvee_{r \in D} \dropp{r}\]

where {\dropp{r}} is the set of packets (in their original form) that match
(and are hence dropped by) rule $r$. These sets are anyway  computed by
our algorithm described above during the normal propagation.
Therefore, to support this
application, we simply save these sets during propagation (see the line
with the comment ``Inferring policy'' in the routine
{\filterruletf{\_}{\_}} in Appendix~\ref{ssec:transf-filt-rule}),
and use them here to
construct the ``drop'' formulas.

In the example in Fig.~\ref{fig:main-example}, the ``accept'' formula for
origin zone Z1 is

\begin{small}
  \begin{center}
  \begin{tabular}{l}
      $[$10.192.29.1-255 : $\neg\{$10.192.29.1-255,
        209.85.153.85, 202.65.23.2$\}]$ \\ 
  \end{tabular}
  \end{center}
\end{small}

which corresponds to $\origpk{\facts{Z2}} \vee \origpk{\facts{Z4}}$. The
``reject'' formula for Z1 is 

\begin{small}
  \begin{center}
  \begin{tabular}{l}
      $[$10.192.29.1-255 : $\{$202.65.23.2, 209.85.153.85$\}]$
  \end{tabular}
  \end{center}
\end{small}

which corresponds to $\dropp{1} \vee \dropp{5}$, where 1 and 5 are rule
numbers in Fig.~\ref{fig:main-example}.

%% Local Variables:
%% tex-main-file: "main.tex"
%% End:

\section{Conclusions, and future work}
\label{sec:concl-future-work}

We presented a novel abstract-interpretation based approach for packet flow
analysis in IP networks. We provided two different variants of the
approach, for inferring different properties, and provided formal claims of
precision of the analysis.  We also illustrated the flexibility of abstract
interpretation in trading precision off for efficiency gains. While we have
taken the first steps in this direction, there are several more-complex
packet-flow analysis settings to which we would like to extend abstract
interpretation. These include (a) accounting for transient changes in
network configuration and topology precisely (transient changes are modeled
by the transitive-closure-based approach of Xie et
al~\cite{xie-infocomm-05}), (b) addressing connection-oriented routing
(i.e., stateful filters), (c) and answering (restricted) forms of temporal
properties of networks. These settings lead to a much larger and richer
state-space than what we have considered in this work. Previous approaches have
not addressed all these issues together; our belief is that abstraction
will be a key ingredient in addressing them with reasonable precision and
scalability. 

%% Local Variables:
%% tex-main-file: "main.tex"
%% End:

%% TODO conclusions and future work

\bibliographystyle{abbrv}
\bibliography{references}

\appendix
\section{Transfer Functions}

\subsection{Transfer function for a filtering rule }
\label{ssec:transf-filt-rule}

\begin{small}
\begin{algorithmic}
  \STATE {\bf Function} \filterruletf{\mathit{rule}:\mbox{a filtering
    rule}}{p \in \absdomain}

\STATE {\bf Output:} ($\mathit{Accepted} \in 2^\absdomain,
\mathit{Unmatched} \in 2^\absdomain$).
\STATE \COMMENT{\emph{Accepted} is a set
  containing zero or one
  abstract
packets, that represent concrete packets represented by $p$ that are accepted
by the rule \emph{rule}. \emph{Unmatched} is a set containing zero or one abstract
packets, that represent concrete packets represented by $p$ that do not match
$\grd{\mathit{rule}}$.}

\STATE $\mathit{Accepted} \leftarrow \phi,
\mathit{Unmatched} \leftarrow \phi$
\IF {$\action{\mathit{rule}}$ == \drop}
	\IF {$\currpk{p}$  $\wedge$ ($\neg \grd{\mathit{rule}}$) $\neq$ \fls}
        \STATE $p1 \leftarrow p$
		\STATE $\currpk{p1}$ $\leftarrow$ $\currpk{p1}$
          $\wedge$ ($\neg$ $\grd{\mathit{rule}}$)
	    \STATE $\origpk{p1}$ $\leftarrow$ $\origpk{p1}$
        $\wedge$ $\neg \mathit{reduce}(\grd{\mathit{rule}},p1)$ \COMMENT {Variant 2} 
        \STATE $\mathit{Unmatched} \leftarrow \{p1\}$
	\ENDIF
	\IF {$\currpk{p}$  $\wedge$ ($\grd{\mathit{rule}}$) $\neq$ \fls}
        \STATE $p2 \leftarrow p$
  	  \STATE $\currpk{p2}$ $\leftarrow$ $\currpk{p2}$ $\wedge$ $\grd{\mathit{rule}}$
	  \STATE $\origpk{p2}$ $\leftarrow$ $\origpk{p2}$
      $\wedge$ $\mathit{reduce}(\grd{\mathit{rule}},p2)$ \COMMENT {Variant 2} 
	  \STATE $\dropp{\mathit{rule}}$ $\leftarrow$
      $\dropp{\mathit{rule}}$ $\cup$ $\{\origpk{p2}\}$ \COMMENT {Inferring policy} 
    \ENDIF
\ELSIF {$\action{\mathit{rule}}$ == ACCEPT}
	\IF {$\currpk{p}$ $\wedge$ $\grd{\mathit{rule}}$ $\neq$ \fls}
        \STATE $p1 \leftarrow p$
		\STATE $\currpk{p1}$ $\leftarrow$ $\currpk{p1}$ $\wedge$ $\grd{\mathit{rule}}$
		\STATE $\origpk{p1}$ $\leftarrow$ $\origpk{p1}$
        $\wedge$ $\mathit{reduce}(\grd{\mathit{rule}},p1)$ \COMMENT {Variant 2} 
		\STATE $\mathit{Accepted}$ $\leftarrow$ $\{p1\}$
	\ENDIF
	\IF {$\currpk{p}$ $\wedge$ ($\neg \grd{\mathit{rule}}$) $\neq$ \fls}
        \STATE $p2 \leftarrow p$
		\STATE $\currpk{p2}$ $\leftarrow$ $\currpk{p2}$ $\wedge$
        ($\neg$ $\grd{\mathit{rule}}$)
		\STATE $\origpk{p2}$ $\leftarrow$ $\origpk{p2}$
        $\wedge$ $\neg \mathit{reduce}(\grd{\mathit{rule}},p2)$ \COMMENT {Variant 2} 
		\STATE $\mathit{Unmatched}$ $\leftarrow$ $\{p2\}$
	\ENDIF
\ENDIF
% \FORALL {$p1$ $\in$ $finalSet\_matched$ \&\& $p2$ $\in$ $finalSet\_unmatched$}
%	\STATE \COMMENT {Inferring policy} $\reach{rule}$ $\leftarrow$ $\reach{rule}$ $\join$ $p1$ $\join$ $p2$
% \ENDFOR
\RETURN (\emph{Accepted, Unmatched})
\end{algorithmic}
\end{small}

\begin{small}
\begin{algorithmic}
  \STATE {\bf Subroutine} \emph{reduce}($g$: a rule's guard, $p \in
  \absdomain$)
  \STATE {\bf Output:} A reduced guard $g'$, which does not refer to NATed
  fields in $p$.
  
  \STATE We assume $g$ to be a conjunction of atomic predicates, each of which
  refers to some field in an abstract packet.
  
  \STATE Let $g'$ be the conjunction of 
  the atomic predicates in $g$ that
  do not refer to any field $i$ such $\ifnated{p}.b_i$ is 1 (this
  conjunction is
  {\tru} if there are no such atomic predicates.)

  \RETURN $g'$
\end{algorithmic}
\end{small}

\subsection {Transfer function for a filtering table}
\label{ssec:transf-filt-table}

\begin{small}
\begin{algorithmic}

\STATE {\bf Function} $\filterchaintf{t:\mbox{a
    a filtering table}}{\mathit{In}: 2^\absdomain}$ 

\STATE {\bf Output:} A set of abstract packets that represent the concrete
packets represented by \emph{In} that are accepted by some rule in  the filtering
table $t$.

\STATE $\mathit{pSet}$ $\leftarrow$ $\mathit{In}$, \emph{Accepted}
  $\leftarrow$ $\phi$

\FORALL {filtering rules $r$ in $t$, in order}
  \STATE $\mathit{pSet'} \leftarrow \phi$
	\FORALL {$p$ $\in$ $\mathit{pSet}$ }		
			\STATE (\emph{Acc, Unmatched)} $\leftarrow$
            \STATE \ \ $\filterruletf{r}{p}$
		
			\STATE $\mathit{pSet'}$ $\leftarrow$ $\mathit{pSet'}$ $\cup$
            \emph{Unmatched}
			\STATE \emph{Accepted} $\leftarrow$ \emph{Accepted} $\cup$ \emph{Acc}
	\ENDFOR
    \STATE $\mathit{pSet}$ $\leftarrow$ $\mathit{pSet'}$
\ENDFOR 

\RETURN \emph{Accepted}
\end{algorithmic}
\end{small}

\subsection {Transfer function for a NATing rule}
\label{ssec:transf-nating-rule}

\begin{small}
\begin{algorithmic}
\STATE {\bf Function} $\natruletf{\mathit{rule}:\mbox{a NATing rule}}{p \in
  \absdomain}$

\STATE {\bf Output:} ($\mathit{Matched} \in 2^\absdomain,
\mathit{Unmatched} \in 2^\absdomain$).
\STATE \COMMENT{\emph{Matched} is a set
  containing zero or one
  abstract
packets, that represent concrete packets represented by $p$ that are matched
by the rule \emph{rule}, and hence will not be passed to subsequent rules
in the chain. \emph{Unmatched} is a set containing zero or one abstract
packets, that represent concrete packets represented by $p$ that do not match
$\grd{\mathit{rule}}$.}

\IF  {\currpk{p} $\wedge$ \grd{rule} $\neq$ \fls}
    \STATE $p1 \leftarrow p$
	\STATE \currpk{p1} $\leftarrow$ \currpk{p1} $\wedge$ \grd{rule}

    \STATE $\origpk{p1}$ $\leftarrow$ $\origpk{p1}$
        $\wedge$ $\mathit{reduce}(\grd{\mathit{rule}},p1)$ \COMMENT {Variant 2} 

    \IF[Variant 2]{$\ifnated{p1}.b_{\natfield{\mathit{rule}}} = 0$}
	\STATE  $p1$ $\leftarrow$
    \updateoriginal{p1}{\mathit{rule}}
    \ENDIF
    
	\STATE  \ifnated{p1}$.b_{\natfield{rule}}$ $\leftarrow$ 1 \COMMENT {Variant 2} 
		
	\STATE $p1$ $\leftarrow$ $\natpacket{p1}{\mathit{rule}}$
				
	\STATE \emph{Matched} $\leftarrow$ $\{p1\}$
\ENDIF
	
\IF  {$\currpk{p}$ $\wedge$ ($\neg$ $\grd{rule}$) $\neq$ \fls}
\STATE $p2 \leftarrow p$
\STATE \currpk{p1} $\leftarrow$ \currpk{p1} $\wedge$ $\neg \grd{rule}$
\STATE $\origpk{p2}$ $\leftarrow$ $\origpk{p2}$ $\wedge$ 
  $\neg \mathit{reduce}(\grd{rule},p2)$ \COMMENT {Variant 2} 
\STATE \emph{Ummatched} $\leftarrow$ $\{p2\}$
\ENDIF

\RETURN (\emph{Matched, Unmatched})
\end{algorithmic}
\end{small}
\subsection{Transfer function for a NATing table}
\label{ssec:transf-nating-table}

\begin{small}
\begin{algorithmic}
\STATE {\bf Function} $\natchaintf{t:\mbox{a
    dnat or snat table}}{\mathit{In}: 2^\absdomain}$ 

\STATE {\bf Output:} A set of abstract packets that represent the concrete
packets represented by \emph{In} after they are transformed by the NATing
rules in $t$.

\STATE $\mathit{pSet}$ $\leftarrow$ $\mathit{In}$, \emph{Out}
  $\leftarrow$ $\phi$

\FORALL {NATing rules $r$ in $t$, in order}
  \STATE $\mathit{pSet'} \leftarrow \phi$
	\FORALL {$p$ $\in$ $\mathit{pSet}$ }		
			\STATE (\emph{Matched, Unmatched)} $\leftarrow$
            \STATE \ \ $\natruletf{r}{p}$
		
			\STATE $\mathit{pSet'}$ $\leftarrow$ $\mathit{pSet'}$ $\cup$
            \emph{Unmatched}
			\STATE \emph{Out} $\leftarrow$ \emph{Out} $\cup$ \emph{Matched}
	\ENDFOR
    \STATE $\mathit{pSet}$ $\leftarrow$ $\mathit{pSet'}$
\ENDFOR 

\RETURN \emph{Out} $\cup$ $\mathit{pSet}$
\end{algorithmic}
\end{small}

\subsection{Transfer function for a Link}
\label{ssec:transf-link}

\begin{small}
\begin{algorithmic}
\STATE {\bf Function} $\forw{(i_1,i_2)}$(\emph{In}: \mbox{a set of abstract
packets})
\STATE {\bf Output:} A set of abstract packets. 

	\STATE $S$ $\leftarrow$ $\natchaintf{\dnat{\nodeof{i_1}}}{\mathit{In}}$
	\STATE $S$ $\leftarrow$ $\filterchaintf{\filter{\nodeof{i_1}}}{S}$
	\STATE $S$ $\leftarrow$ $\natchaintf{\snat{\nodeof{i_1}}}{S}$
    \STATE Construct a filtering table $t$ with a single filtering
    rule $r$ that drops all packets that don't
    satisfy formula $\route{\nodeof{i_1}}(i_1)$.
    \STATE $S \leftarrow \filterchaintf{t}{S}$
    \RETURN $\bigjoin S$
\end{algorithmic}
\end{small}

\subsection{Transfer functions for \emph{natPacket} and\\
  \emph{update\_original\_packet} functions}

\begin{small}
\begin{algorithmic}
  \STATE {\bf Subroutine} $\natpacket{p \in \absdomain}{r:\mbox{a NATing rule}}$
  \STATE {\bf Output:} A copy of packet $p$ in which the field
  $\natfield{r}$ of $\currpk{p}$
  has been overwritten with the range of values
  $\action{r}$. 
\end{algorithmic}

\begin{algorithmic}
  \STATE {\bf Subroutine}\\ $\updateoriginal{p \in \absdomain}%
   {r:\mbox{a NATing rule}}$
  \STATE {\bf Output:} A copy of abstract packet $p$,
  in which the field $\natfield{r}$ of $\origpk{p}$ has been overwritten
  with the contents of field $\natfield{r}$ of $\currpk{p}$. 
\end{algorithmic}
\end{small}

We do not provide a formal definition of the above two formulas, which need
to simulate updation of fields by formula manipulation. Rather, we provide
an illustration of how they work. 
Let's consider an example where the packet header has two fields with 2 bits
to represent each field.

Let $p$ be the given packet and $\currpk{p}$ be defined by the formula: 

$ ((b_1 \wedge \neg b_2) \vee (b_1 \wedge b_2)) \wedge \neg b_3 \wedge b_4$ \\

where $b_1, b_2, b_3$, and $b_4$ are the four bits in the header of
$\currpk{p}$. 

Now let the given NATing rule be $r$ and $\grd{r}$ be \tru. Let the action
of $r$ be to change the value of first field in packet header to 00.
Then the new value of $\currpk{p}$, computed by $\natpacket{\_}{\_}$ is
given by the formula:

$ (((b_1 \wedge \neg b_2) \vee (b_1 \wedge b_2)) \wedge \neg b_3\wedge b_4)  \wedge$\\
$ ((\neg b_1') \wedge (\neg b_2') \wedge (b_3' = b_3) \wedge (b_4' = b_4))$

wherein the primed variables are now treated as the free variables. The
first line in the formula above represents the original value of
$\currpk{p}$, while the second line captures the fact that while bits $b_1$
and $b_2$ are both set to 0 bits $b_3$ and $b_4$ are preserved.

Say before we do the NATing above
the \ifnated{p} field in the packet $p$ is 01,
indicating that field 1 has not been NATed previously and field 2 has
been NATed previously. Also, 
let $\origpk{p}$ be defined by the formula:

$(\neg c_3 \wedge \neg c_4)$

where $c_1, c_2, c_3$, and $c_4$ are the four bits in the header of
$\origpk{p}$. 
Since we are now updating the first two bits of the packet $\origpk{p}$, and these have
not been NATed before, we update $\origpk{p}$ to the following formula:

$ (((b_1 \wedge \neg b_2) \vee (b_1 \wedge b_2)) \wedge \neg b_3 \wedge b_4$) $\wedge$ \\
$(\neg c_3 \wedge \neg c_4)$ $\wedge$\\
$((c_1' = b_1) \wedge (c_2' = b_2) \wedge (c_3' = c_3) \wedge (c_4' = c_4))$
  
wherein the primed variables are to be treated as the free variables.
The first line above represents the old value of $\currpk{p}$; the second
line represents the old value of $\origpk{p}$; the last line indicates that
the new value of bits $c_1$ and $c_2$ are to the same as the values of old
values of the bits $b_1$ and $b_2$, respectively. This captures the fact
that first two bits of $\currpk{p}$ are being copied to $\origpk{p}$. 

We can simplify both formulas generated above by first eliminating unprimed
variables, and then renaming the primed variables to their corresponding
unprimed form. 

%\subsection { Transfer function for routing rule}
% For each incoming packet and each routing rule forward the packet on the interface where the guard matches the packet

%\begin{algorithmic}
%\STATE Let $In\_p$ be the incoming packet.

%\IF {$In\_p.curr$ $\wedge$ $rule.guard$ $\neq$ \fls}
%		\STATE $In\_p.curr$ $\leftarrow$ $In\_p.curr$ $\wedge$ $rule.guard$
%		\RETURN $In\_p.curr$
%\ELSE
%		\RETURN % Macro for NULL packet
%\ENDIF
%\end{algorithmic}

%\subsection {Transfer function for routing chain}

%\begin{algorithmic}

%\STATE Let $In\_p$ be the incoming packet.
%\STATE $p$ $\leftarrow$ $In\_p$

%\FORALL {$rule$ $\in$ $Routing Rule$}
%	\STATE $p$ $\leftarrow$ $\routeruletf{p}{rule}$
%\ENDFOR
%\end{algorithmic}

\end{document}